\documentclass[noshowpacs,superscriptaddress,groupedaddress,aps]{revtex4}  
\usepackage{graphicx}  
\usepackage{dcolumn}   
\usepackage{bm}        
\usepackage{amssymb}   
\usepackage{xcolor}
\usepackage{amsmath,amssymb,graphics,graphicx,subfigure}

\begin{document}

\title{Quantum key distribution with entangled photons generated on-demand by a quantum dot}

\author{Francesco Basso Basset}
\thanks{These authors contributed equally.}
\affiliation{Department of Physics, Sapienza University of Rome, 00185 Rome, Italy}
\author{Mauro Valeri}
\thanks{These authors contributed equally.}
\affiliation{Department of Physics, Sapienza University of Rome, 00185 Rome, Italy}
\author{Emanuele Roccia}
\affiliation{Department of Physics, Sapienza University of Rome, 00185 Rome, Italy}
\author{Valerio Muredda}
\affiliation{Department of Physics, Sapienza University of Rome, 00185 Rome, Italy}
\author{Davide Poderini}
\affiliation{Department of Physics, Sapienza University of Rome, 00185 Rome, Italy}
\author{Julia Neuwirth}
\affiliation{Department of Physics, Sapienza University of Rome, 00185 Rome, Italy}
\author{Nicol\`{o} Spagnolo}
\affiliation{Department of Physics, Sapienza University of Rome, 00185 Rome, Italy}
\author{Michele B. Rota}
\affiliation{Department of Physics, Sapienza University of Rome, 00185 Rome, Italy}
\author{Gonzalo Carvacho}
\affiliation{Department of Physics, Sapienza University of Rome, 00185 Rome, Italy}
\author{Fabio Sciarrino}
\thanks{e-mail: fabio.sciarrino@uniroma1.it; rinaldo.trotta@uniroma1.it}
\affiliation{Department of Physics, Sapienza University of Rome, 00185 Rome, Italy}
\author{Rinaldo Trotta}
\thanks{e-mail: fabio.sciarrino@uniroma1.it; rinaldo.trotta@uniroma1.it}
\affiliation{Department of Physics, Sapienza University of Rome, 00185 Rome, Italy}

\date{\today}

\begin{abstract}
Quantum key distribution---exchanging a random secret key relying on a quantum mechanical resource \cite{Scarani2009}---is the core feature of secure quantum networks. Entanglement-based protocols offer additional layers of security \cite{Waks2002,Acn2007} and scale favorably with quantum repeaters \cite{Guha2015}, but the stringent requirements set on the photon source have made their use situational so far. Semiconductor-based quantum emitters are a promising solution in this scenario, ensuring on-demand generation of near-unity-fidelity entangled photons \cite{Huber2018} with record-low multi-photon emission \cite{Schweickert2018}, the latter feature countering some of the best eavesdropping attacks \cite{Ltkenhaus2002, Duek1999}. Here we first employ a quantum dot to experimentally demonstrate a modified Ekert quantum key distribution protocol \cite{Acn2006} with two quantum channel approaches: both a $250$ meter long single mode fiber and in free-space, connecting two buildings within the campus of Sapienza University in Rome. Our field study highlights that quantum-dot entangled-photon sources are ready to go beyond laboratory experiments, thus opening the way to real-life quantum communication.
\end{abstract}

\maketitle

The quantum mechanical properties of single photons have proven to be fundamental for the implementation of intrinsically secure cryptographic distribution systems and are the stepping stone for the development of quantum networks \cite{Flamini2018}. One of the technologically ambitious tasks sought after within such approach is to consistently reduce the multi-photon emission using single-photon emitters without suffering from a trade-off in the photon generation rate \cite{Takeoka2015}. The reason is mainly attributed to different attack strategies, such as number splitting \cite{Ltkenhaus2002} or beam splitting attacks \cite{Duek1999}, taking advantage on the non-perfect single photon distribution of the emitter and hacking the security of the protocol. Sub-Poisson light with low $g^{(2)}(0)$ can improve communication security in presence of channel losses \cite{Waks2002c, Waks2002b}. On-demand photon emitters offer a good solution to these issues at the hardware level \cite{Allaume2004} in order to make the photon distribution nearly unassailable. Furthermore, the exploitation of EPR non-locality \cite{Bennett1992, Ekert1991} establishes another crucial tool for the prevention of key errors, and a subsequent improvement on security of the communication procedure against individual attacks \cite{ Waks2002}. Finally, under certain experimental requirements, it could also provide device-independent operation \cite{Ma2007, Acn2007}. Semiconductor quantum dots (QDs) are a promising platform for the accomplishment of all these tasks, due to their low multi-photon emission rate \cite{Schweickert2018}, increasing brightness \cite{Liu2019,Wang2019} and on-demand production of high-purity entangled states \cite{Huber2018}. Nonetheless, their application has focused so far on single-photon prepare-and-measure protocols \cite{Waks2002b,Takemoto2015}, with only a single in-lab test of the BBM92 protocol displaying near-failure baseline error and insufficient throughput \cite{Dzurnak2015}. Moreover, in none of these works entangled photons have been generated on demand, one of the key features distinguishing QDs from standard sources based on parametric down-conversion.

\begin{figure*}
\centering{\includegraphics[width=0.95\textwidth]{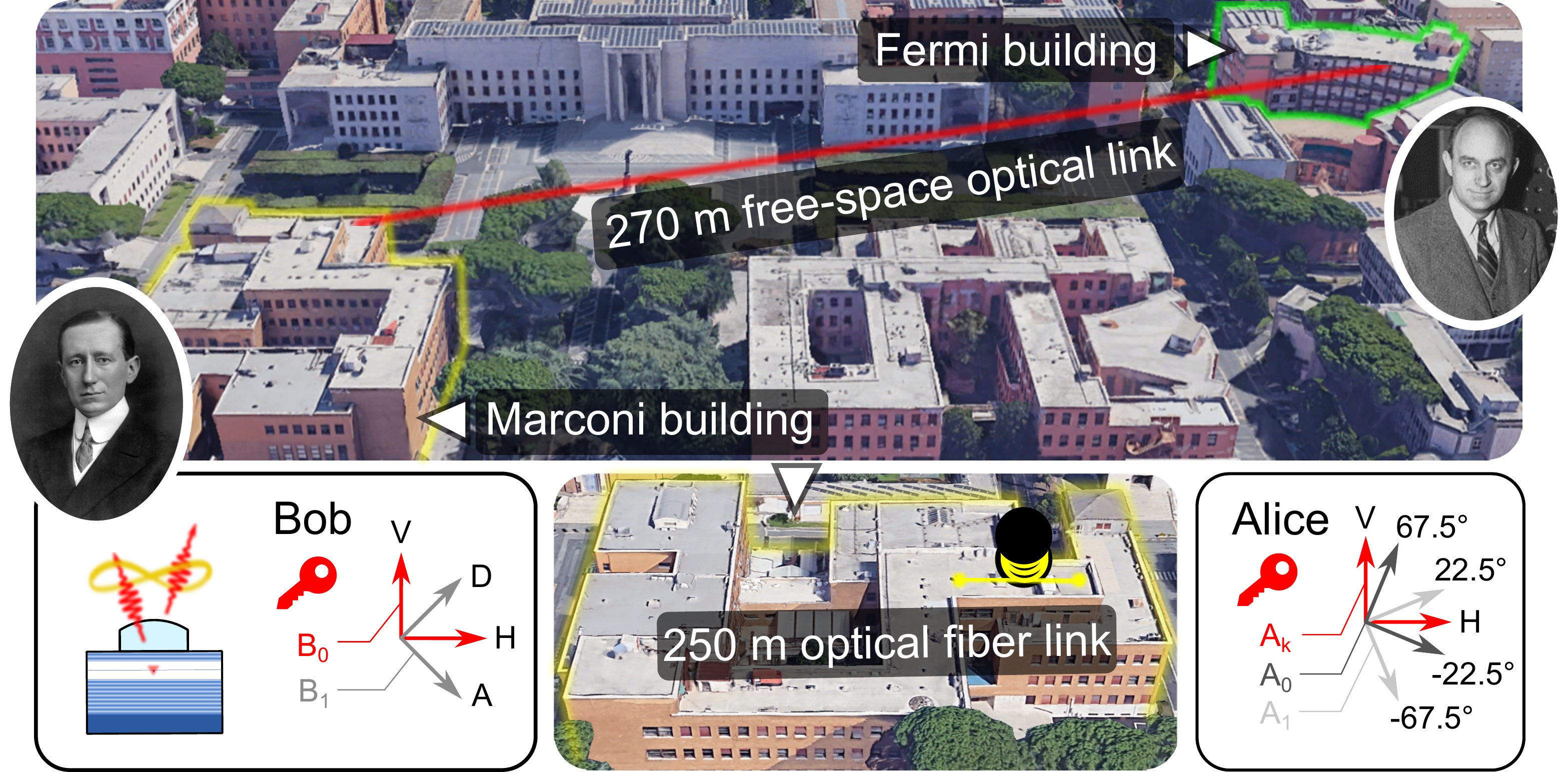}}
\caption{\textbf{Entanglement-based QKD and optical links overview.} Entangled photon pairs generated by a single QD are shared across the Sapienza University campus in Rome over a $270$ m free-space distance and, in addition, between two laboratories in the same building via a $250$ m SMF. Map data: Google Earth. We illustrate the main concepts of the asymmetrical Ekert approach: after traveling through the optical link connecting the users, the photons are measured by Alice and Bob on the measurement bases $\lbrace A_k$, $A_0$, $A_1 \rbrace$ and  $\lbrace B_0$, $B_1\rbrace$ in the Fermi and Marconi building respectively. In this case, the combination of the horizontal and vertical polarization states $\lbrace A_k,B_0 \rbrace$ constitute the basis to share the secure key. In parallel, the other pairs ensure verification of the entanglement quality, by measuring the Bell parameter of the two-photon state.}
\label{fig:conceptual}
\end{figure*}

In this letter, we present the first experimental implementation of an entanglement-based quantum key distribution (QKD) protocol---specifically the asymmetrical Ekert proposal \cite{Acn2006, Ekert1991}---with the use of entangled photons generated by a nearly deterministic QD-based source. In particular, we demonstrate the viability of our QKD system under different choices of quantum channel. Figure~\ref{fig:conceptual} gives an overview of our study, which recreates two urban communication scenarios within the campus of Sapienza University of Rome, namely via a single-mode fiber (SMF) and through free-space covering a distance of approximately $270$ m. This double approach is motivated by the fact that, on the one hand, networks based on fiber communication are the common solution within urban environments, due to their scalability with moderate losses for short distances. On the other hand, over long distances free-space links still represent the best choice to connect users due to its low signal attenuation \cite{Bedington2017, Yin2020} and the possibility of sending complex states such as those exploiting the orbital angular momentum (OAM) of light---something still under development with optical fibers---despite the need for more complex sender and receiver systems.

\begin{figure*}[ht]
\centering
{\includegraphics[width=0.9\textwidth]{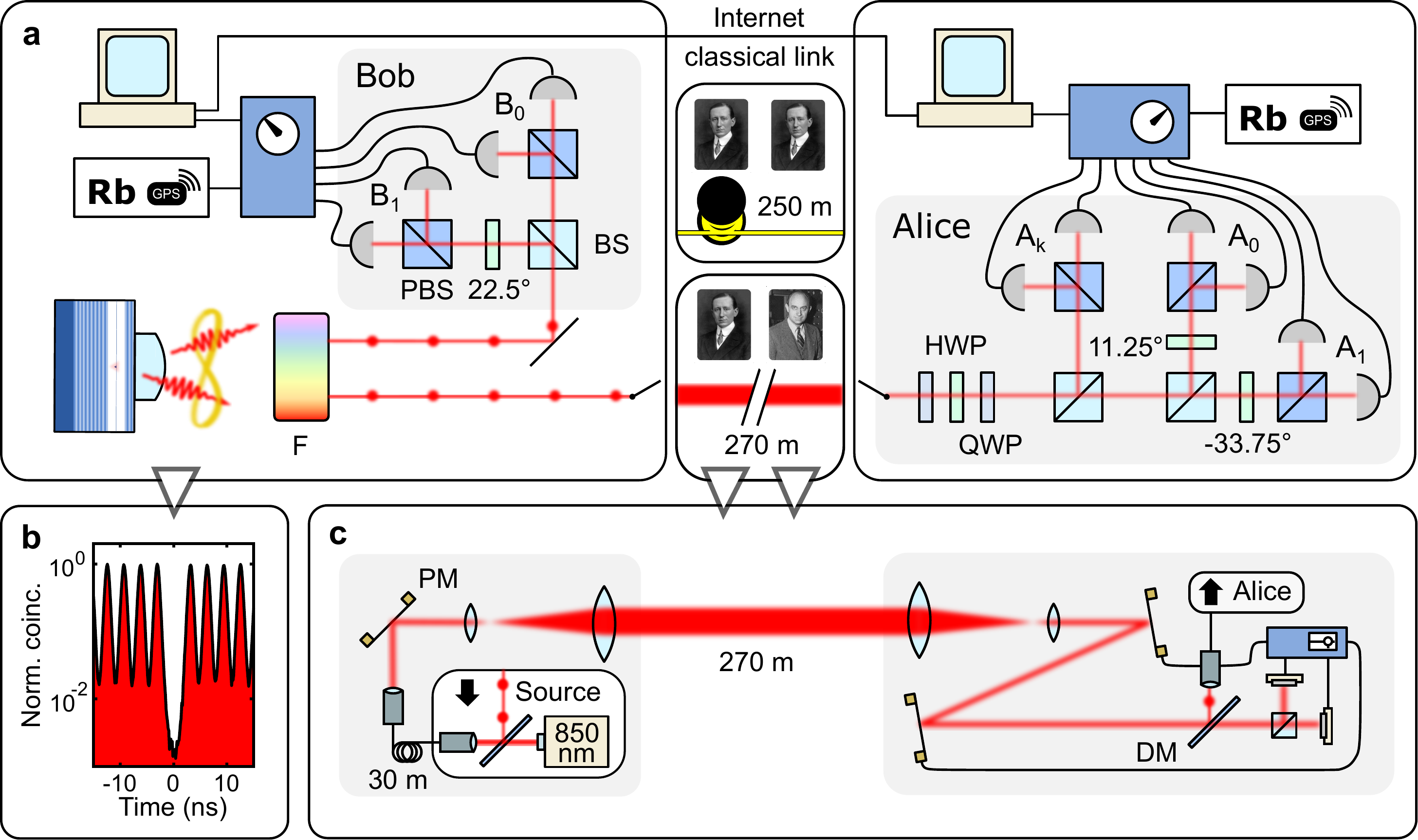}}
\caption{\textbf{Experimental realization of the QKD protocol. a} Illustration of the setup used for the entanglement-based QKD protocol. On Bob side, the single GaAs QD generates two entangled photons that are separated by spectral filtering (F). One photon goes directly to the Bob measurement apparatus, while the other travels through one of the quantum channels depicted in Fig.~\ref{fig:conceptual}. The photons arriving to Alice's station are compensated in polarization with a set of two quarter-wave plates (QWPs) and a half-wave plate (HWP). After a random splitting with 50:50 beam splitters (BS)---which also mirror the polarization of the reflected beam---the polarization states are measured in the bases $\lbrace B_0$, $B_1\rbrace$ on Bob and $\lbrace A_k$, $A_0$, $A_1 \rbrace$ on Alice using HWPs and polarizing beam splitters (PBS). The photons are finally collected and detected by avalanche photodiodes connected to two independent time taggers. These are synchronized combining a GPS signal and Rb oscillators. \textbf{b} Autocorrelation histogram of the X emission line, showing the low multi-photon component of the source. \textbf{c} Sender and receiver in free-space communication. A diode laser beam at $850$ nm is sent together with the single photon signal and feeds a closed stabilization system, which controls the piezoelectric mirror mounts (PM). The QD signal is selected by a dichroic mirror (DM) and then coupled into a SMF directed to Alice's measurement apparatus.}
\label{fig:setup}
\end{figure*}

Figure~\ref{fig:conceptual} illustrates the principle of operation of the QKD protocol. The two parties randomly select a measurement to perform on their subsystem of an entangled state from a set of linear polarization bases. In a quarter of the cases, Alice and Bob pick a combination of the $\lbrace A_k,B_0 \rbrace$ bases, and their local reference frames are aligned in such a way that they will get the same result out of the measurement. When Alice and Bob share among themselves the information that they performed the same measurement, its random outcome is a bit added to the shared secret key. In presence of noise or imperfect entangled states, the keys may differ by an amount quantified by the quantum bit error rate (QBER),
\begin{equation}
QBER= (1-E(A_k,B_0))/2,
\end{equation}
where $E$ is the correlation coefficient, i.e., the expectation value on the $\lbrace A_k,B_0 \rbrace$ pair of measurements. 

In the meanwhile, the security of the QKD is monitored via an entanglement measurement on other acquisition channels. The measurement bases on Alice ($\lbrace A_0,A_1 \rbrace$) and Bob ($\lbrace B_0,B_1 \rbrace$) sides are chosen in order to obtain the maximum value of the Bell parameter $S$, checking the violation of Bell inequality $|S|\leq 2$, accordingly to the CHSH figure of merit \cite{Clauser1969},
\begin{equation}
S=E(A_0,B_0)+E(A_0,B_1)+E(A_1,B_0)-E(A_1,B_1).
\end{equation} 

Such procedure is a convenient variation of the well-known Ekert proposal. The asymmetrical scheme reduces the number of required detectors with respect to the original Ekert91 protocol, as reported in the practical implementation illustrated in Fig.~\ref{fig:setup}a. At the same time, the fraction of photons dedicated to the key exchange is increased, while the security check is still performed by monitoring the Bell inequalities \cite{Acn2006}. Additionally, the scheme has been demonstrated viable for device-independent operation \cite{Acn2007}. 

In our experiment Bob is placed near the entangled photon source, while Alice is on the other side of the employed quantum channel. The mentioned measurements on Alice ($\lbrace A_k$, $A_0$, $A_1 \rbrace$) and Bob ($\lbrace B_0$, $B_1 \rbrace$) sides are associated to detection events on different avalanche photodiodes. The coincidences between Alice and Bob are obtained by exploiting a Rb clock system, which, combined with a GPS, allows synchronization at nanosecond time scale.

The photon pairs shared among the two parties are deterministically emitted in the maximally entangled state $|\phi^+ \rangle = 1/\sqrt{2} (|\text{HH}\rangle + |\text{VV}\rangle)$ from a single GaAs QD, pumped with a $320$ MHz repetition rate laser. The key transmission along the two different quantum channels was performed using two QDs with very similar features (see full comparison in Methods). The entanglement fidelity reaches up to $94.1\mbox{--}95.8\%$ due to the choice of QDs with sub-$\mu$eV fine structure splitting (FSS). This ensures a high polarization indistinguishability between the two photons generated by the radiative biexciton-exciton (XX-X) cascade. The low multi-photon emission efficiency is checked by measuring the auto-correlation function $g^{(2)}(|\tau| < 0.8~ns)$, normalized by the side peaks from consecutive laser pulses. Figure~\ref{fig:setup}b reports an auto-correlation measurement run on the exciton emission line in laboratory conditions, corresponding to $g^{(2)}_X(|\tau| < 0.8~ns) = 0.0034(2)$, a value mainly limited by the detector time resolution. With a similar procedure we measure the cross-correlation function between the two photons of the entangled pair and find a fidelity of preparation of $94.3(3)\%$, due to the deterministic resonant two-photon excitation of the QD \cite{Jayakumar2013}. This demonstrates near on-demand generation of entangled photons. 

The main difference between the two protocol realizations resides in the additional setup required for the signal transmission over free-space, which is shown in Fig.~\ref{fig:setup}c. As in this work we explore both fiber-based and free-space QKD, we emphasize that the latter has proven significantly more challenging, as it requires a different level of complexity from the technical standpoint, especially for the measures needed to counteract atmospheric turbulence. The system we set up is composed of two matching telescopes and an active-stabilization system, which is guided by tracking a reference 850 nm laser on two position-sensitive detectors. The system is designed to counteract the effect of beam wander created during the propagation hence allowing a single-mode coupling of the photon signal (at 785 nm) with approximately 40\% efficiency.

\begin{figure*}[ht]
\centering
{\includegraphics[width=0.95\textwidth]{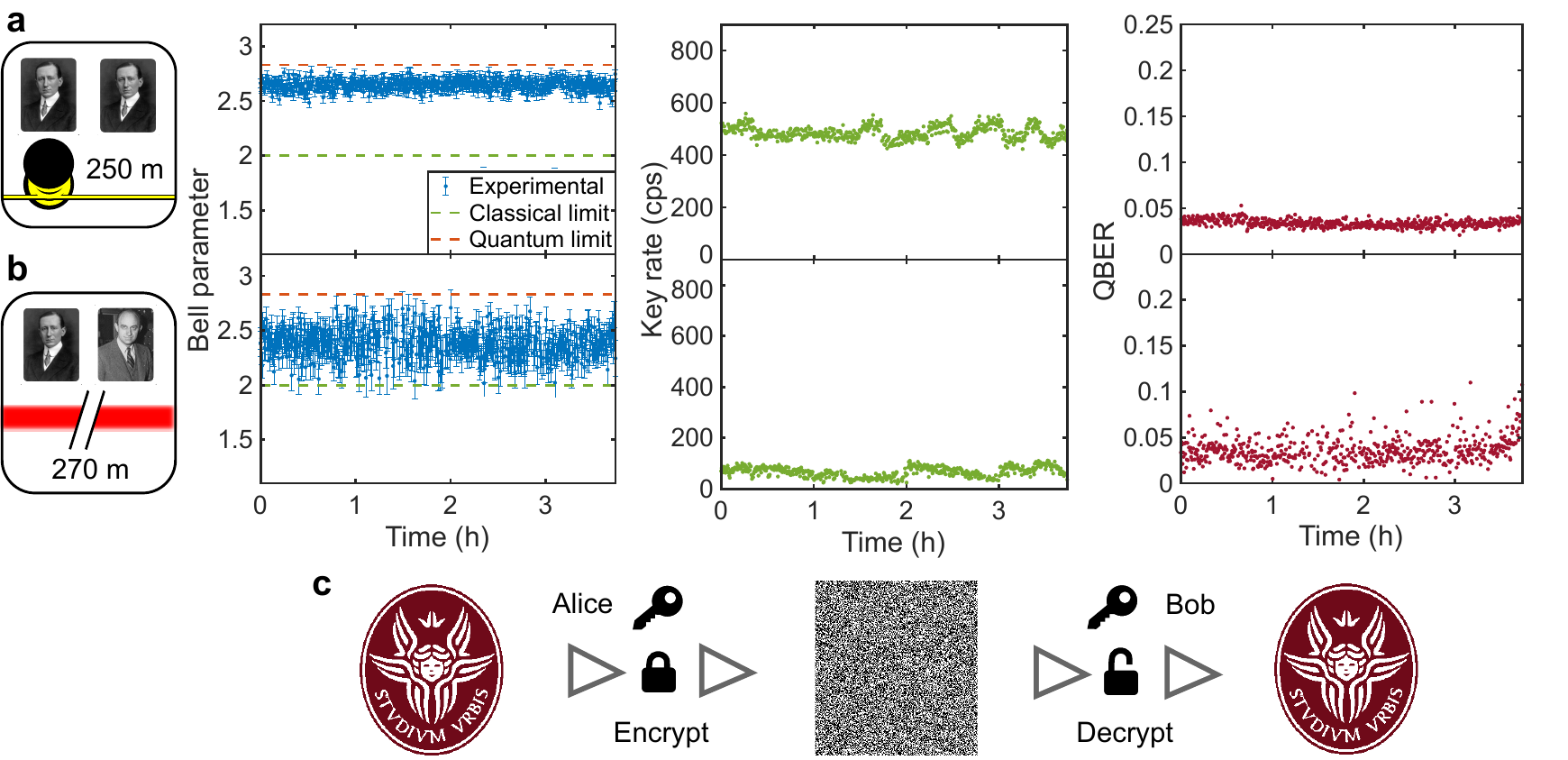}}
\caption{\textbf{Experimental key sharing via the modified Ekert protocol. a, b} The Bell parameter, raw key rate and QBER measured with both the SMF optical link, panel \textbf{a}, and the free-space optical link, panel \textbf{b}. In the latter case, the data transmission was conducted overnight and interrupted at dawn. Each data point corresponds to $5$ acquisitions of $1.2$ s. The error bars in the Bell parameter are calculated by Gaussian propagation assuming a Poissonian distribution for the coincidence counts. \textbf{c} The encryption and decryption of Sapienza logo using the one-time pad technique with the shared free-space key, after the error correction and randomness enhancement steps.}
\label{results}
\end{figure*}

The comparison of the experimental results between the fiber and free-space approaches is illustrated in Fig.~\ref{results}. The data is synchronously collected on the two sides in packets of $1.2$ s acquisition time and shared over the university network, for a total time of $224$ minutes. For both optical links, we measure the QBER, the Bell parameter $S$ and the amount of key shared among the two parties. Using the fiber communication approach, in Fig.~\ref{results}a, a total $217.76$ kB key is shared, with a mean key rate of $486$ bit s$^{-1}$. Rates as high as $785$ bit s$^{-1}$ were obtained in preliminary tests using a common time-to-digital converter for the two parties and a different QD from the same sample (see Supplementary Information). The mean QBER of the key is $Q_{SMF} = 0.0337(2)$, while the mean Bell parameter is $S_{SMF} = 2.647(2)$.

In free-space communication, see Fig.~\ref{results}b, we manage to share a $34.589$ kB-long key string, relying on $60$ bit s$^{-1}$ of mean key rate. In this case, the average values of QBER and Bell parameter are $Q_{FS} = 0.040(2)$ and $S_{FS} = 2.37(10)$ respectively.

These results are the very first field demonstration of successful QKD using a QD-based deterministic source of entangled photons. Both optical link choices showed a substantial violation of Bell inequality, demonstrating the two-photon entanglement preservation over the quantum communication channel. The reliability of the key quality shared during the communication is verified in both cases by monitoring the QBER, which remains consistently well below the critical insecure value of $11\%$. Part of the distributed key was used to encrypt and decrypt the Sapienza logo as depicted in Fig.~\ref{results}c. After the binary error correction process \cite{Colombier2017}, we apply a Trevisan extractor to restore uniform randomness \cite{Mauerer2012} and compensate for setup-induced polarization changes.

The comparison of the measurements highlights some interesting considerations on the quantum channel choice performing an EPR-based QKD approach using QDs. Atmospheric turbulence and the complex stability requirements of optical apparatuses for free-space communication lead to signal loss and performance degradation when compared with the in-fiber solution for short distances, which is unavoidable to a certain degree, despite the room for optimization. However, by increasing the distance between the QKD users, the free-space communication is expected to deteriorate less dramatically, becoming a more advisable solution. Moreover, it allows to use the orbital angular momentum degree of freedom for multiplexing or independence from local reference frames \cite{Vallone2014}.

In conclusion, we have experimentally performed an entanglement-based QKD protocol with the use of a QD photon-pair source, demonstrated to be viable both with in-fiber and free-space quantum communication channels. Exactly 20 years after the first theoretical proposal on the possibility of using quantum dots to generate regulated and entangled photons \cite{Benson2000}, our study demonstrates that this semiconductor-based quantum technology is mature to go out of the lab, and further improvements---discussed in detail in the Supplementary Information---will soon open the way to real-life quantum communication. In particular, we envisage that the possibility of interfacing entangled photons from QDs with the same or other quantum systems \cite{Trotta2016}, together with the prospect of enhancing photon extraction \cite{Liu2019, Wang2019}, will be the key to boost secure quantum cryptography over large distances.

\section*{Methods}
\subsection*{Entangled photon source}
Polarization-entangled photons are generated by a single GaAs QD embedded in a crystalline matrix of Al$_{0.4}$Ga$_{0.6}$As. The QDs are fabricated using the Al droplet etching technique, as described in reference \cite{BassoBasset2019}. Due to the presence of distributed Bragg reflectors in the sample structure and to the use of a hemispherical solid immersion lens, an extraction efficiency of approximately $8\%$ is achieved in the investigated sample. This value allows to employ the source in realistic quantum communication schemes, but further improvements are required to overcome state-of-the-art SPDC, as quantitatively discussed in the Supplementary Information. Two different QDs from the same sample and with similar characteristics were used in the two parts of the experiment. The multi-photon emission is very similar, $g^{(2)}_X(|\tau| < 0.8~ns) = 0.0034(2)$ and $g^{(2)}_{XX}(|\tau| < 0.8~ns) = 0.0041(3)$ for the QD used in the fiber experiment, $g^{(2)}_X(|\tau| < 0.8~ns) = 0.0040(4)$ and $g^{(2)}_{XX}(|\tau| <~0.8 ns) = 0.0045(4)$ for the QD used in the free-space experiment. Achieving these values does not require polarization suppression for the cancellation of background radiation from the laser. The entanglement fidelity is only slightly different, $95.8(1.2)\%$ in the free-space case and $94.1(1.0)\%$ in the in-fiber case, due to the different fine structure splitting, $0.35$ and $0.85$  $\mu$eV respectively. The single photon count at the output of the first SMF, disregarding losses in the quantum channels and in the Alice and Bob apparatuses, is $700$ and $620$ kcps for the free-space and in-fiber QD respectively. 
\subsection*{Setup description}
The entangled photon source is kept at $5$ K in a low-vibration closed cycle He cryostat. The optical excitation is performed with a Ti:Sapphire laser together with a $4$f pulse shaper, to reduce its bandwidth to $0.1$ nm, and two Mach–Zehnder interferometers, to increase its repetition rate four-fold. The signal is collected from the QD using a $0.81$ NA objective and the backscattered laser light is filtered out with notch filters. The two photons from a single XX-X cascade are then separated by two volume Bragg gratings, collected by two SMFs and distributed to the measurement setups of the two QKD users, Alice and Bob. The SMF quantum channel consists of a 780-HP fiber with 80\% transmission after its 250 m of length at the wavelength of operation (785 nm).
In the free-space experiment, a $850$ nm diode laser is collected through a $30$ m SMF together with the photon associated to the exciton line and brought to the transmission platform. Here, the beam is magnified by a factor of $6$ by exploiting a telescope, with the aim of keeping collimation and reducing the effect of beam wandering during the $270$ m travel in air, where the atmospheric attenuation losses amount at $10\%$. A mirror with piezoelectric adjusters is used to compensate for slow drifts in the pointing direction. 
On the receiver side, the beam diameter is reduced with a telescope similar to the one used by the sender. This permits to couple the signal in a SMF connected to Alice's apparatus. The $850$ nm laser is separated from the QD signal using a dichroic mirror, and sent to two position-sensitive detectors that provide feedback for an active beam stabilization system implemented with two mirrors with piezoelectric adjusters. The single-photon signal is finally collected in the SMF with an average coupling efficiency of $40\%$ and sent to Alice's measurement apparatus. A more detailed account of the stabilization strategy and the channel losses is presented in the Supplementary Information.
The QD signal is finally collected by single photon-silicon avalanche diodes. The detection events are recorded on each side by a time-digital converter which is also connected to a GPS-locked rubidium oscillator, acting as a reference clock. A more in-depth description of the synchronization procedure is found in the Supplementary Information.

\bibliographystyle{naturemag}
\bibliography{QKD_bib}

\section*{Acknowledgments}
This work was financially supported by the European Research Council (ERC) under the European Union’s Horizon 2020 Research and Innovation Programme (SPQRel, grant agreement no. 679183), by the Regione Lazio program “Progetti di Gruppi di ricerca” legge Regionale n. 13/2008 (SINFONIA project, prot. n. 85-2017-15200) via LazioInnova spa, by MIUR (Ministero dell’Istruzione, dell’Università e della Ricerca) via project PRIN 2017 “Taming complexity via QUantum Strategies a Hybrid Integrated Photonic approach” (QUSHIP) Id. 2017SRNBRK. We thank Armando Rastelli and Saimon F. Covre da Silva for providing high-quality GaAs QD samples, Paolo Mataloni for very fruitful discussions, Antonio Miriametro for helpful technical assistance.
\section*{Authors’ contributions}
The experiment was performed, in the Marconi building, by Francesco Basso Basset, Emanuele Roccia, Julia Neuwirth and, in the Fermi building, by Mauro Valeri, Gonzalo Carvacho, Davide Poderini and Nicolò Spagnolo. Davide Poderini and Emanuele Roccia wrote the software for data acquisition and secret key extraction.  Francesco Basso Basset and Emanuele Roccia performed the source characterization. Mauro Valeri, Valerio Muredda, and Gonzalo Carvacho designed and assembled the receiver setup and the active stabilization system. Francesco Basso Basset, Emanuele Roccia and Michele B. Rota designed and assembled the source and sender setup. Francesco Basso Basset, Emanuele Roccia, Mauro Valeri, Davide Poderini wrote the manuscript with feedback from all authors. All the authors participated in the discussion of the results. Rinaldo Trotta and Fabio Sciarrino conceived the experiments and coordinated the project.
\section*{Competing interests}
The authors declare no competing financial interests.

\end{document}